*Research paper*
# Explicit cloud representation in the *Atmos* 1D climate model for Earth and rocky planet applications.

**Thomas Fauchez**[1,2,5], **Giada Arney**[1,4,5], **Ravi Kumar Kopparapu**[1,3,4,5], **Shawn Domagal Goldman**[1,4,5]

[1] NASA Goddard Space Flight Center, 8800 Greenbelt Road, Greenbelt, MD 20770, USA
[2] Goddard Earth Sciences Technology and Research (GESTAR), Universities Space Research Association, Columbia, MD 21046, USA
[3] Department of Astronomy, University of Maryland College Park, College Park, MD 20742, USA
[4] NASA Astrobiology Institute's Virtual Planetary Laboratory, P.O. Box 351580, Seattle, WA 98195, USA
[5] Sellers Exoplanet Environments Collaboration, NASA Goddard Space Flight Center

* **Correspondence:** thomas.j.fauchez@nasa.gov

**Abstract:**

1D climate models are less sophisticated than 3D global circulation models (GCMs), however their computational time is much less expensive, allowing a large number of runs in a short period of time to explore a wide parameter space. Exploring parameter space is particularly important for predicting the observable properties of exoplanets, for which few parameters are known with certainty. Therefore, 1D climate models are still very useful tools for planetary studies. In most of these 1D models, clouds are not physically represented in the atmosphere, despite having a well-known, significant impact on a planetary radiative budget. This impact is simulated by artificially raising surface albedo, in order to reproduce the observed-averaged surface temperature (i.e. 288 K for modern Earth) and a radiative balance at the top of the atmosphere. This non-physical representation of clouds, causes atmospheric longwave and shortwaves fluxes to not match observational data. Additionally, this technique represents a parameter that is highly-tuned to modern Earth's climate, and may not be appropriate for planets that deviate from modern Earth's climate conditions. In this paper, we present an update to the climate model within the *Atmos* 1D atmospheric modeling package with a physical representation of clouds. We show that this physical representation of clouds in the atmosphere allows both longwave and shortwave fluxes to match observational data. This improvement will allow us to study the energy fluxes for a variety of cloudy rocky planets, and increase our confidence in future simulations of temperature profile and net energy balance.

---

## 1. Introduction

Clouds are main actors on a planetary radiative budget (see (1) for Earth; (2) for Venus; (3) for Titan). Clouds may reflect a significant fraction of incoming star light back to space, leading to a cooling of the atmosphere via this albedo effect. On the other hand, clouds may absorb outgoing thermal infrared radiation emitted by the planetary surface, leading to a warming of the atmosphere. Which of these opposite cloud radiative effects is dominant depends on the cloud type, composition, and altitude, the planetary surface temperature and stellar spectrum. Note that clouds are normally approximated to be "grey" absorbers because their absorption is mostly independent of wavelength, compared to gases that absorb specific regions of the spectrum that correspond to specific molecular vibrational–rotational modes.

While clouds are well known to be very important actor of climate, some 1-dimensional (altitude) climate models of early Earth and other planets are cloud-free and instead adjust the surface albedo to reflect the cloud albedo and to reproduce the mean surface temperature (4). Although this assumption seems crude, it has been used widely in numerous 1D climate models ((5-9); (10, 11), (12),(13),(14),(15, 16)). In 1D climate models, the atmosphere is represented in a single, horizontally homogeneous vertical column. 3D planetary climate models – such as ROCKE3D (17), LMD (18), CAM (19), UM (20), etc. – include clouds explicitly. However, the 1D models have much shorter runtimes allowing the exploration of a wider parameter space, include additional modules (e.g., for complex photochemistry; (15, 16)), and the simulation of planets over geological timescales of billions of years.

(21) tested the validity of the assumption that clouds' effects can be simulated by tuning surface albedo, and have shown that mean energy budgets deviate far from observations for both shortwave and longwave radiations. That paper also demonstrated that a 1D climate model can reproduce the observed global mean energy budget of Earth if clouds are explicitly parameterized in the atmosphere according to observed cloud climatologies. In this research letter, we use the (21) methodology to explicitly represent clouds in the 1D climate model contained in the *Atmos* 1D modeling suite, and therefore to better represent cloud effects on future atmospheric studies with *Atmos*. Because Atmos also contains a flexible photochemical model that can simulate the chemical profiles of rocky planets for a wide variety of boundary conditions, this will improve our ability to incorporate the effect of clouds, chemistry, and radiative transfer in a single, flexible atmospheric package. The manuscript is organized as follows: the materials and methods are described in section 2, the comparison of our results with current Earth mean energy budget is given in section 3 and conclusions and outlooks are highlighted in section 4.

## 2. Materials and Methods

*2.1. Overview*

We used the *Atmos* 1D, radiative–convective, cloud-free climate model. This is an open-sourced model derived from the one historically used by the Kasting group, most recently updated to simulate habitable zones around stars (14). The model uses correlated-k absorption coefficients derived from the HITRAN 2008 and HITEMP 2010 databases for pressures of $10^{-5}$-$10^2$ bar and for temperatures of 100– 600 K. For Earth simulation, we consider a 1976 US standard atmosphere with the height ranging from 0 to 80 km (~$10^{-5}$) and modern-day greenhouse (GH) gases of 369 ppmv carbon dioxide ($CO_2$), 1790 ppbv methane ($CH_4$), and 316 ppbv nitrous oxide ($N_2O$) included a $H_2O$ water profile. Oxygen ($O_2$) and ozone ($O_3$) are also set to present day concentrations. Similarly, present solar flux is used with a zenith angle of 60°, which is the value that reproduces global annual climate properties. *Atmos* is utilized to simulate the climates of other rocky planets, including conditions proposed for ancient Venus, Earth, and Mars, and rocky exoplanets. The model has historically been used to simulated planets within the habitable zone.

*2.2. Cloud representation*

*2.2.1. Climatology*

Clouds are very difficult to represent in GCMs and particularly the convection processes. They are one of the largest sources of differences and uncertainties between GCMs (IPCC AR5 report (22)). Also, cloud climatology is complicated to measure. For passive, or even active sensors, clouds are opaque after a certain optical depth through the cloud, due to a finite penetration depth that depends on the wavelength of the observation and sensor properties. Therefore, for the same portion of the sky, observing either from the surface or from the top of the atmosphere (TOA) may lead to different cloud statistics and resulting climatologies.

Acknowledging this limitation, (23) showed that the vertical distribution of clouds in the Earth atmosphere can be resolved in three distinct cloud layers. (24) have estimated their zonally-averaged cloud fraction profiles using a combination of radiosonde and satellite data (from the International Satellite Cloud Climatology Program (ISCCP)). According to these observational data, and following the modeling strategy introduced by (21), we have prescribed three cloud layers:
- **Low clouds** corresponding to liquid water clouds such as cumulus, stratocumulus and stratus
- **Mid clouds** corresponding to altocumulus, altostratus and nimbostratus
- **High clouds** corresponding to ice clouds, i.e. cirrus, cirrocumulus and cirrostratus

Details about these cloud layers are given in Table 1. We can see that each cloud is defined by its thermodynamic phase (ice or liquid), top and base pressures and altitudes, water path, effective size of cloud particle, and cloud fraction. The average cloud altitudes/pressures have been obtained from (24), and the effective radius of cloud particles were obtained from (21).

*2.2.2. Cloud optical properties*

Cloud optical properties are often represented by three parameters: the extinction coefficient ($\sigma$ in $km^{-1}$), the single scattering albedo ($\varpi_0$), and the asymmetry factor of the scattering phase function ($g$). Liquid cloud droplets optical properties are calculated from Mie theory. The current version of the code allows the use of cloud droplet effective radii from 1 to 50 μm with an increment of 1 µm for a log-normal distribution, but is very flexible to any change of effective radius, size distribution, etc.

Contrary to the Mie theory, there is no exact solution of the radiative transfer inside ice crystals. Also, ice cloud optical properties are difficult to characterize because of the diversity of crystal sizes, shapes and orientations. Several parametrizations were developed for visible and infrared wavelengths ((25), (26-28), (29, 30), (31-34)).

*Atmos* allows for two parameterizations for ice crystal optical properties:
- An aggregate ice crystal model (28) with a monodisperse distribution. This is the model we have selected in this study with an effective radius of 48.70 µm (see Table 1) to match the generalized effective size used in (21).
- An ensemble model of complex ice particle shapes (33, 34) based on *in situ* measurements of more than 20,000 particle-size distributions (35), given the optical coefficients as a function of IWC and temperature. the particularity of these optical properties is that they allow for a rigorously consistent representation of clouds in the shortwave and long-wave (36). Also, the use of this parameterization could be useful to predict optical coefficients from the water content and temperature.

Note that we have evaluated the differences in term of fluxes and surface temperature when the Yang or Baran ice crystal optical property is used. The difference is only of the order of ~1 W/m² for the longwave fluxes. Shortwave fluxes and the surface temperature are not affected.

### 2.2.3. Water path

Climatological values for the water path are difficult to obtain because most of the measurements are made from satellites which are not always able to distinguish the water path contribution between overlapping cloud layers. Therefore, in order to match (21) simulations we chose the same water paths of 20, 25 and 32 g·m$^{-2}$ for high, mid and low clouds, respectively, close to climalogical data of (23).

**Table 1. Cloud characteristics used in this study following (23, 24) and (21).**

| Cloud type | High | Mid | Low |
|---|---|---|---|
| **Phase** | Ice | Liquid | Liquid |
| **Base pressure (bar)** | 0.30 | 0.60 | 0.85 |
| **Top pressure (bar)** | 0.25 | 0.50 | 0.70 |
| **Base altitude (m)** | 8028 | 3668 | 1018 |
| **Top altitude (m)** | 9020 | 5086 | 2309 |
| **Water Path (g·m$^{-2}$)** | 20 | 25 | 32 |
| **Effective radius (µm)** | 48.70 | 11 | 11 |
| **Cloud fraction** | 0.25 | 0.25 | 0.40 |

### 2.2.4. Cloud fraction

The main challenge to account for prescribed clouds in a 1D climate model resides in the parameterization of the cloud fraction. The mean global cloud fraction (the cloud fraction regardless the cloud type) has high temporal and spatial variation. (23) have estimated the mean global cloud fraction of 67.6%. Three combinations of the three cloud layers are possible in climate models and GCMs: 1) random overlap; 2) maximum overlap; or 3) a combination of both. In our study, we assume that cloud layers are randomly overlapped and that each homogeneous cloud layer can be represented by a single fraction from which the overlap is calculated. Indeed, in our model each cloud layer is separated by intervals of clear sky leading to no correlation between cloud layers (*e.g.* (37) and (21)). When assuming random overlap, the cloud fractions of each layers which satisfy the ICCP mean global cloud fraction are: $f_{high}$=0.25, $f_{mid}$=0.29 and $f_{low}$=0.39, with $f_{high}$, $f_{mid}$, and $f_{low}$ representing the cloud fraction of high, mid and low altitude clouds, respectively. In order to compare our modeling results with those of (21) we use their cloud fractions of $f_{high}$=0.25, $f_{mid}$=0.25 and $f_{low}$=0.40.

Because *Atmos* is a 1D climate model with a single, horizontally homogeneous, vertical column it is not possible to performed a heterogeneous cloud coverage simulation in a single run. Instead, we have done one run per permutations ($2^N$=8 permutations for N=3 clouds layers) and weighted their outputs (longwave and shortwave fluxes, temperature profile, mixing ratio etc.) by their combined cloud fraction. While many configurations are possible to obtain the value of the global mean cloud fraction and individual cloud fractions, we use and show only one of them in Table 2. The mean global combined cloud fraction (MGCCF) being the sum of the cloud fraction of each ($2^N$-1) cloudy permutation and the clear sky cloud fraction CSCF=(1-MGCCF). Using this method, clouds are physically included in the radiative transfer code, instead of tuning the surface albedo (4).

**Table 2: Cloud overlap categories and corresponding combined cloud fractions using ICCP data (24).**

| **Cloud overlap** | Clear | High | Mid | Low | High + Mid | High + Low | Mid + Low | High + Mid + Low | Mean global |
|---|---|---|---|---|---|---|---|---|---|
| **Combined fraction** | 0.34 | 0.14 | 0.09 | 0.24 | 0.03 | 0.03 | 0.08 | 0.05 | 0.66 |

*2.3. Radiative transfer*

The radiative transfer in *Atmos* is computed using the correlated-K method, with 8-term k-coefficients for H2O and CO2. A δ two-stream approximation (38) is used to calculate the net absorbed solar radiation for each of the 101 layers. Longwave and shortwave radiation are treated in two different subroutines.

Each layer of the atmosphere is represented by its optical properties, namely the extinction coefficient ($\sigma$ in km$^{-1}$) or optical thickness $\tau$, the single scattering albedo ($\varpi_0$), and the asymmetry factor of the scattering phase function (*g*). In a single atmospheric layer, multiple components can contribute to the optical properties: the gas (including Rayleigh scattering) and aerosol extinction were previously included; here we add cloud extinction. While multiple sources of extinctions are present in an atmospheric layer (assumed to be horizontally homogeneous), the code can compute only one $\sigma$, $\varpi_0$ and *g*. To obtain the resulting optical coefficients a weighting method is applied:

- The combined extinction $\sigma^{tot}$ (gas + Rayleigh + aerosols + cloud) is simply the sum of each extinction, with the total optical thickness of the atmospheric layer $\tau^{tot} = \sigma^{tot} \times dZ$ where *dZ* is the geometrical thickness of the layer in km.

- The combined single scattering albedo $\varpi_0$ between aerosols, Rayleigh and cloud is calculated as: $\varpi_0^{tot} = (\varpi_0^{rlt} \times \tau^{rlt} + \varpi_0^{aer} \times \tau^{aer} + \varpi_0^{cld} \times \tau^{cld})/(\tau^{tot})$ with $\varpi_0^{tot}$, $\varpi_0^{rlt}$, $\varpi_0^{aer}$ and $\varpi_0^{cld}$ the total, Rayleigh, aerosol and cloud single scattering albedos, respectively and $\tau^{rlt}$, $\tau^{aer}$ and $\tau^{cld}$ the Rayleigh, aerosol and cloud optical thicknesses.
- The combined asymmetry parameter $g^{tot} = (\tau^{rlt} \times \varpi_0^{rlt} \times g^{rlt} + \tau^{aer} \times \varpi_0^{aer} \times g^{aer} + \tau^{cld} \times \varpi_0^{cld} \times g^{cld})/(\tau^{rlt} \times \varpi_0^{rlt} + \tau^{aer} \times \varpi_0^{aer} + \tau^{cld} \times \varpi_0^{cld})$.

With the updated version of *Atmos* presented in this paper, the cloud radiative effect is easily solved in each layer where a cloud is present.

## 3. Results

We have compared modern Earth's global mean annual budgets between measurements ((39),(40)), the simulation of (21), and the results of our simulation (*Fig. 1*). Shortwave fluxes – ~342 W·m$^{-2}$ at the top of the atmosphere (TOA) – are either absorbed or scattered by the clouds and the atmosphere, or pass through to the surface, which can also reflect them according to the surface albedo, which in these simulations are set to a value of 0.125, based on global measurements (39). Longwave radiation is mostly emitted from the surface, and can be partly absorbed by the atmosphere and/or absorbed and scattered by clouds. The difference between the incoming TOA shortwave flux and the outgoing shortwave flux is equal to the outgoing longwave flux at TOA when the system is at radiative equilibrium. Both the prior work by (21) and our simulations demonstrate radiative equilibrium., to within 0.2 W·m$^{-2}$.

The comparison plots in Fig. 1 indicate that our model agrees well with the measured global mean annual budgets for the modern Earth and (21). In particular, they are all in agreement with a surface temperature of ~288 K and similar fluxes emitted, transmitted, absorbed or scattered by the various components of the atmosphere. Therefore, this figure illustrates that clouds have been correctly prescribed in our model.

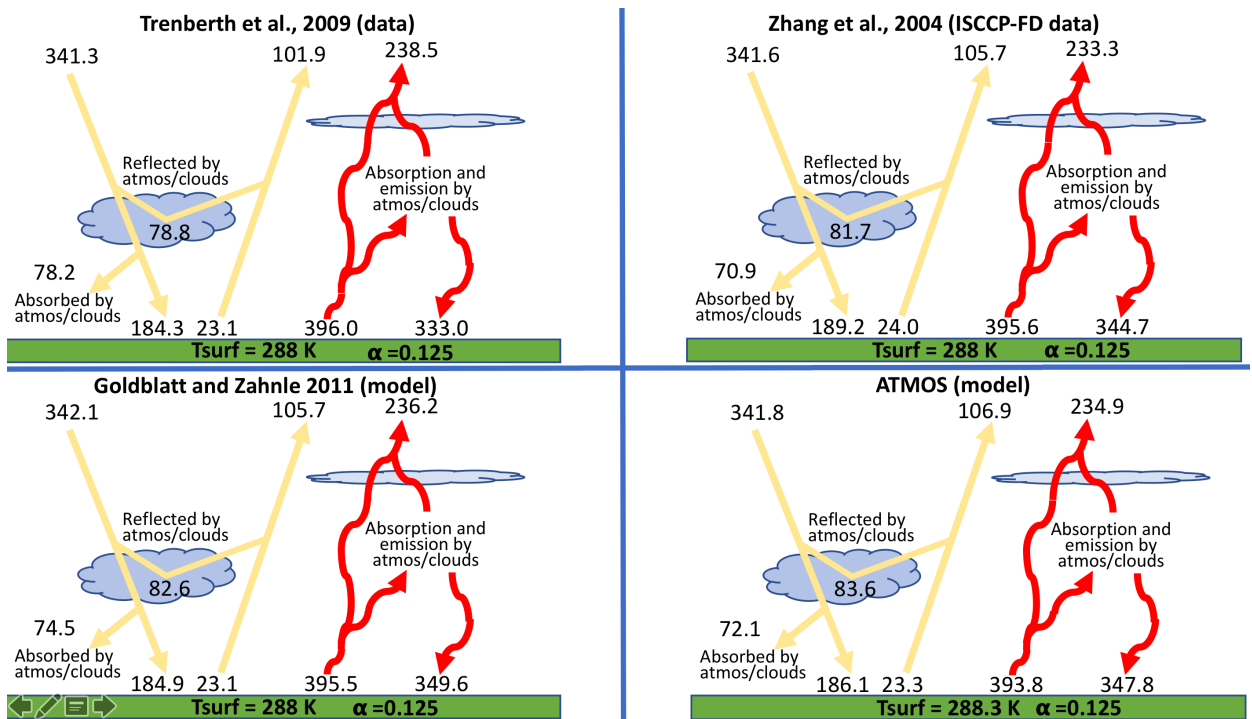

Figure 1: Global annual mean energy budgets. top left: composite data from (39), top right: ISCCP-FD data from (40), bottom left: (21) simulation and bottom right: simulation from this study. α represents the surface albedo.

We also compare these global annual budgets using *Atmos* without clouds (using the (4) Kasting's approximation) and with clouds physically prescribed in the atmosphere for modern Earth and Archean Earth insolation and atmospheric composition (*Fig. 2*). Details on the atmospheric composition of the Archean Earth case are given in Table 2. Both models achieve radiative equilibrium, but in cloud-free model, the surface albedo is tuned to 0.32, in order to achieve TOA energy balance. This is significantly higher than the Earth's measured surface albedo, of 0.125 (39), which is the value used in our simulations with clouds.

Concerning the modern Earth case (uper row of Fig. 2), while the radiative balance at TOA and a 288 K surface temperature have been obtained in the cloud-free and cloudy model, the longwave and shortwave fluxes are very different. Much more shortwave flux is going down to the surface in the case without clouds, and combined with a larger surface albedo much more flux is reflected by the surface. However, because no clouds are physically present in the atmosphere, only a small contribution of the Rayleigh scattering is added to the upward shorwave flux. There are less dramatic changes for the longwave fluxes, except that in the case without clouds, there is about $\sim$+20 W.m$^{-2}$ at TOA and $\sim$-20 W.m$^{-2}$ down to the surface. Overall, the physical representation of clouds in the atmosphere significantly change the atmospheric fluxes and allows a much better agreement with observation data (see Fig. 1).

Similar results are observed for the Archean Earth case (lower row of Fig. 2). However because the insolation is 20% lower than modern Earth value and that the greenhouse gas concentration is larger, this reduces the albedo effect and the relative greenhouse effect of clouds, respectively, leading to smaller differences of shortwave and longwave fluxes between the cloud-free and cloudy models.

*Table 3: Parameters used in the Archean Earth simulation at 1 bar surface pressure.*

| Solar cnst. | $N_2$ | $CO_2$ | $CH_4$ | $O_2$ | $H_2$ | Ar |
|---|---|---|---|---|---|---|
| 0.8 | 9.6640E-01 | 2.0000E-02 | 3.4169E-03 | 9.7626E-09 | 1.5064E-04 | 9.7626E-03 |

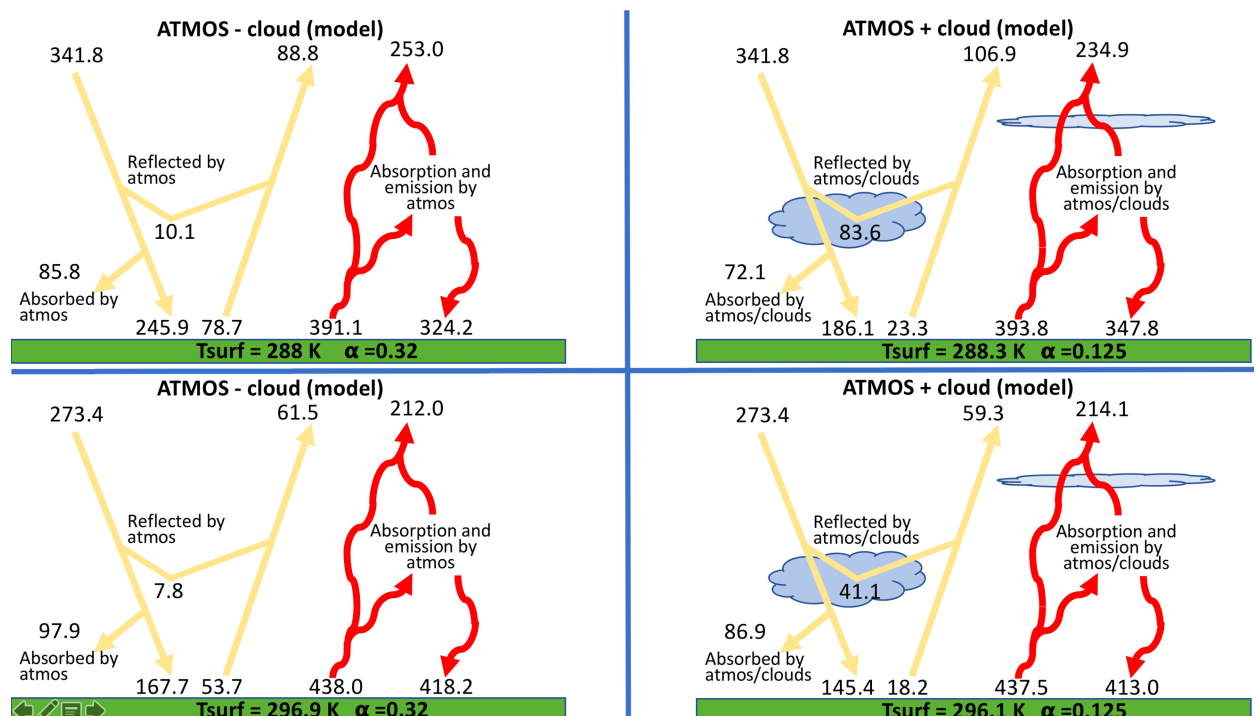

**Figure 2:** Comparison of the cloud-free (left column) and cloudy (right column) model for Modern Earth (top row) and Archean Earth (bottom row) insolation and atmospheric composition.

We finally ran a sensitivity study between the cloud-free and cloudy *Atmos* version at various insolation (0.95, 1.0 and 1.05 present day value) and various $CO_2$ partial pressures (0.5, 1.0, 2.0, 3.0 and 4.0 present day value). Results are shown in Fig. 3 which represents the radiative forcing at TOA (top panel) estimated from the difference between the outgoing infrared flux ($F_{IR}$) and the modern Earth TOA outgoing infrared flux $F_{IR}^0$, as well as the surface temperature difference ($T_{surf}$ - $T_{surf}^0$) estimated from the surface temperature ($T_{surf}$) and the modern Earth surface temperature ($T_{surf}^0$), as a function of the relative $CO_2$ partial pressure ($pCO_2 / pCO_2^0$) and solar constant S. The cloud coverage is assumed constant (modern Earth value).

We can see in Fig. 3 that the cloud-free model overestimates and underestimates the radiative forcing and surface temperature when the insolation S is increased or decreased, respectively. Indeed, as we can also see in Fig. 4 showing the planetary albedo for the cloud-free and cloudy models, at various insolation and $pCO_2 / pCO_2^0$, the planetary albedo is larger for the cloudy model. in the cloud-free case, an increase of the insolation will directly impact the surface (minus the atmospheric absorption/reflection), while in the cloudy case, the clouds albedo mitigate the increase of solar flux received on the surface, producing less warming. The opposite is also true when the insolation is reduced, the cloud greenhouse effect keeping the atmosphere warmer than the cloud-free model.

Also, at a given insolation, the cloud-free model is more sensitive to changes of the $CO_2$ partial pressure $pCO_2$ (from 0.5 to 4.0 times the modern Earth value of 360 ppm). Indeed, clouds also produce a greenhouse effect and are optically thick in the water vapor windows, mitigating the impact of the change of $pCO_2$ with respect to the cloud-free model. Without clouds, the atmosphere is more transparent (especially in the water vapor windows) and the change of $pCO_2$ have a large effect. Note that at the lowest insolation (S=0.95) the cloud-free and cloudy models tend to converge to a similar TOA forcing and surface temperature with increasing $pCO_2$ because i) the albedo effect of clouds is reduced at low insolation and ii) the higher the pCO2 the lower the cloud contribution to the greenhouse effect. Clouds therefore act as climate dampers, and a cloud-free model would systematically overestimate the forcing by greenhouse gases or insolation changes.

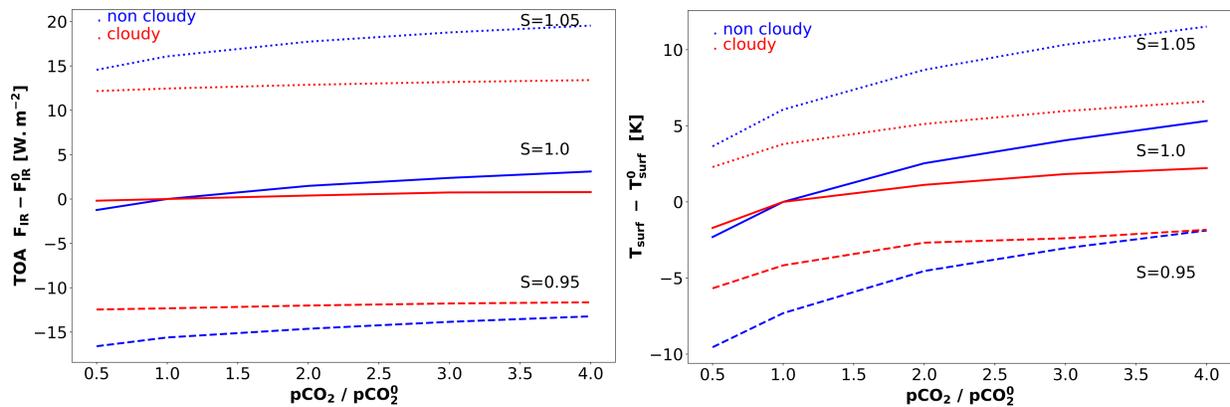

**Figure 3: TOA outgoing infrared flux relative to the modern Earth TOA outgoing infrared flux ($F_{IR}$ - $F_{IR}^0$, left panel) surface temperature relative to modern Earth surface temperature ($T_{surf}$ - $T_{surf}^0$, right panel) as a function of the relative $CO_2$ partial pressure ($pCO_2$ / $pCO_2^0$) and solar constant S.**

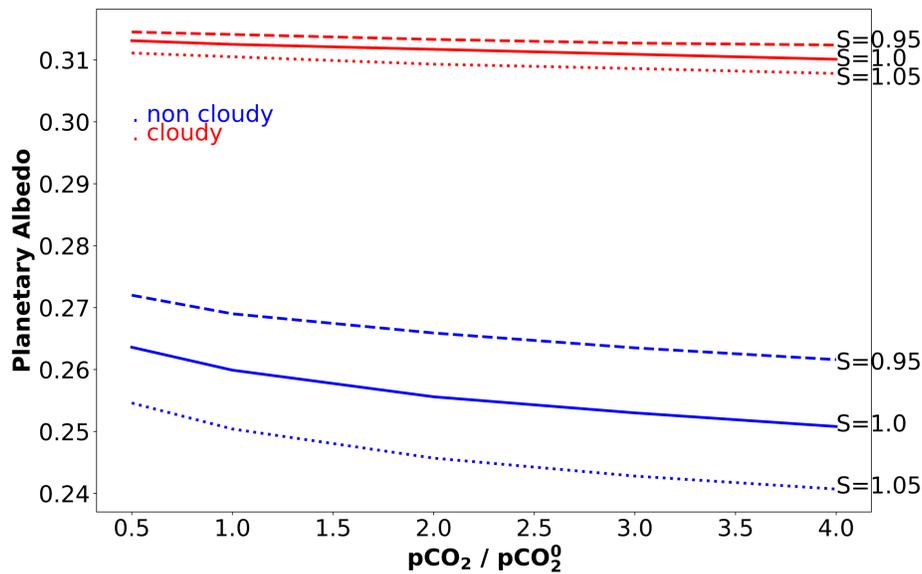

**Figure 4: Planetary albedo for the cloud-free (blue) and cloud (red) model at various insolation S and partial pressure of $CO_2$ relative to modern Earth value ($pCO_2/pCO_2^0$).**

## 4. Conclusions and outlook

The aim of this paper is twofold: 1) to present an update of the *Atmos* 1D, radiative–convective, cloud-free climate model (14), with clouds physically prescribed in the atmosphere; and 2) to validate this approach with a limited set of simulations. This introduces a publicly availably tool that impoved the representation the cloud radiative effect on both atmospheric longwave and shortwave fluxes, compared to earlier versions of the model. We use a similar methodology to the one developed by (21) but adapted it to the specifics of the *Atmos* code.

The physical representation of clouds in the radiative transfer of *Atmos* allows us to accurately match the atmospheric fluxes of observational data of the Earth radiative budget. However, such simulation is eight times more computational expensive because the code needs to be run once per cloud permutation. Note that we have limited the number of cloud layers to three in this study, according to the definition of the ICCP, but up to five layers can be use in this version (more can be easily included if needed).

The implications of this work are broader than simulations of Earth's climate. This approach can be applied to a variety of terrestrial planets, including paleo Earth, ancient Venus, ancient Mars or rocky exoplanets. However, an important caveat is that this introduces additional "knobs" into the model. Determining the right values for those knobs for planets without observational cloud data (like we have for modern Earth) is a challenge. This is something that could be overcome by single GCM simulations for a similar part of parameter space, or parameterization of the cloud cover values to bound potential climate and habitability outcomes. Such parameterizations would not require additional computational time, as they only require the simulations of the 8 cloud-cover cases. Future work will explore the validity of such approaches, with comparisons to existing GCM simulations for planets other than modern-day Earth.

The comparison study with archean Earth (Fig. 2), and the sensitivity study varying the solar insolation and $CO_2$ partial pressure (Fig. 3), illustrate the versatality of our code for various solar fluxes and atomspheric compositions. Shortwave and longwave fluxes as well as surface and atmospheric temperature profile are key parameters to evaluate the inner and outer edges of the habitable zone and they would be better represented by 1D cloudy climate models than 1D cloud-free models previously used. Furthermore, the photochemical and climate models in the *Atmos* package are capable of simulating the vertical profiles of chemistry, temperature, and photochemical aerosols. This paper demonstrates an ability to add the effects of clouds to simulations with the *Atmos* package. This means we can now simulate the global effects of clouds and simulate surface climate in a more self-consistent manner. It will have implications for future reasearch on the ancient climates of Earth and Mars, simulations of long-term climate change on rocky worlds, and on studies of the habitable zones of other stars.

Acknowledgments: R. K gratefully acknowledges funding from NASA Habitable Worlds grant NNX16AB61G. G.A, R.K and S.D.D-G also acknowledge funding from NASA Astrobiology Institute's Virtual Planetary Laboratory lead team, supported by NASA under cooperative agreement NNH05ZDA001C. The authors are thankful for support from GSFC Sellers Exoplanet Environments Collaboration (SEEC), which is funded by the NASA Planetary Science Divisions Internal Scientist Funding Model.